\documentclass[aps,prb,preprint,showpacs,superscriptaddress]{revtex4-1}

\usepackage{graphicx}
\usepackage{amssymb}
\usepackage{amsmath}
\usepackage{bm}

\begin{document}

\title{Manipulation of Magnetic Solitons on Odd-Numbered Macrospin Rings}

\author{J. P. Morgan}
\affiliation{Department of Physics, University of Cambridge, Cambridge CB3 1HE, United Kingdom}
\affiliation{School of Physics \& Astronomy, University of Leeds, Leeds LS2 9JT, United Kingdom}

\author{R. P. Cowburn}
\affiliation{Department of Physics, University of Cambridge, Cambridge CB3 1HE, United Kingdom}

\date{\today}

\begin{abstract}
We report simulations of a frustrated odd-numbered macrospin ring system, with full point dipolar interactions, driven by a rotating uniform applied magnetic field of constant magnitude. The system is designed with equally-spaced radially-aligned macrospins, which must carry a frustrated soliton defect in its ground state. It is shown how correctly tuning the applied field magnitude can allow for non-trivial unidirectional propagation of the soliton, the required directional pressure acquired via the curvature of the ring. Furthermore, the system, which may be employed as a multiple rotation counter, is tested for robustness against quenched disorder as would be present in an experimental realization.
\end{abstract}


\maketitle

\section{Introduction}
The conception and construction of systems of well-defined coupled macrospins underpins both the fields of artificial frustrated magnetism\cite{2006Wang} and nanomagnetic logic.\cite{2006Imre,2011Niemier} The two communities however remain largely separate, balanced at the fundamental and applied ends of the same physical problem: the predictable and controlled evolution of magnetic configurations in patterns of nanomagnets. A carefully-designed balance of field scales allows for the manipulation of well-defined defects or ``frustrations'' in the local ground state (GS) macrospin order, forming the basis of interesting and useful operations. Quenched disorder (QD), the distribution in properties between the coupled components inherent from nanopatterning, however acts to disrupt these processes.\cite{2011Niemier,2012BudrikisMorgan}

Antiferromagnetic Ising lattices\cite{1996Davidovic} and ice models\cite{2006Wang,2008Qi} have been realized from patterned elements possessing well-defined bi-stable dipolar behavior, in which competing interactions control collective ordering. Propagation of charge defects has generated substantial interest\cite{2010Budrikis,2010Ladak,2011Morgan2,2011Mengotti,2012Phatak} due to a qualitative analogy with ``monopole'' excitations in rare-earth pyrochlore materials.\cite{2008Castelnovo,2009Mol} Magnetic islands and multilayer heterostructures have also been employed for information processing, in the form of logic gates, shift registers and ratchets.\cite{2000Cowburn2,2002Allwood,2006Imre,2012FernandezPacheco,2013Lavrijsen,2013Bhowmik,2013Kiermaier,2014Breitkreutz}
A domain wall ``soliton''\cite{1994SolitonsBook} at the boundary between two GS ordered phases can be unidirectionally field-driven along a conduit given underlying symmetries are appropriately broken.

In this work, we introduce a novel system which exemplifies the equality of such contemporary works in nanomagnetism, and explore its potential in executing reliable and repeatable operations. The system is a circular ring of radially-aligned evenly-spaced Ising-like spin moments. Crucially, the number of moments $n$ is fixed to be odd, which, as we will show, forces the system to  possess a frustrated topological soliton defect\cite{1994SolitonsBook,2000Cowburn2} in its GS and form an approximate realization of a magnetic M\"{o}bius loop.\cite{2004Cador}
The curvature of the ring imposes chirality under the application of a rotating constant-amplitude magnetic field, and we use numerical simulations to show how this allows for a soliton to be driven around the system. With experimental realization in mind, we make various assumptions appropriate for patterned nanomagnets to build the model and further test for robustness against QD. Furthermore, we discuss the application of the system as a multiturn counter.\cite{2006Mattheis,2007Diegel,2009Diegel,2012Mattheis}

\begin{figure}
\begin{center}
\includegraphics[width=1\columnwidth]{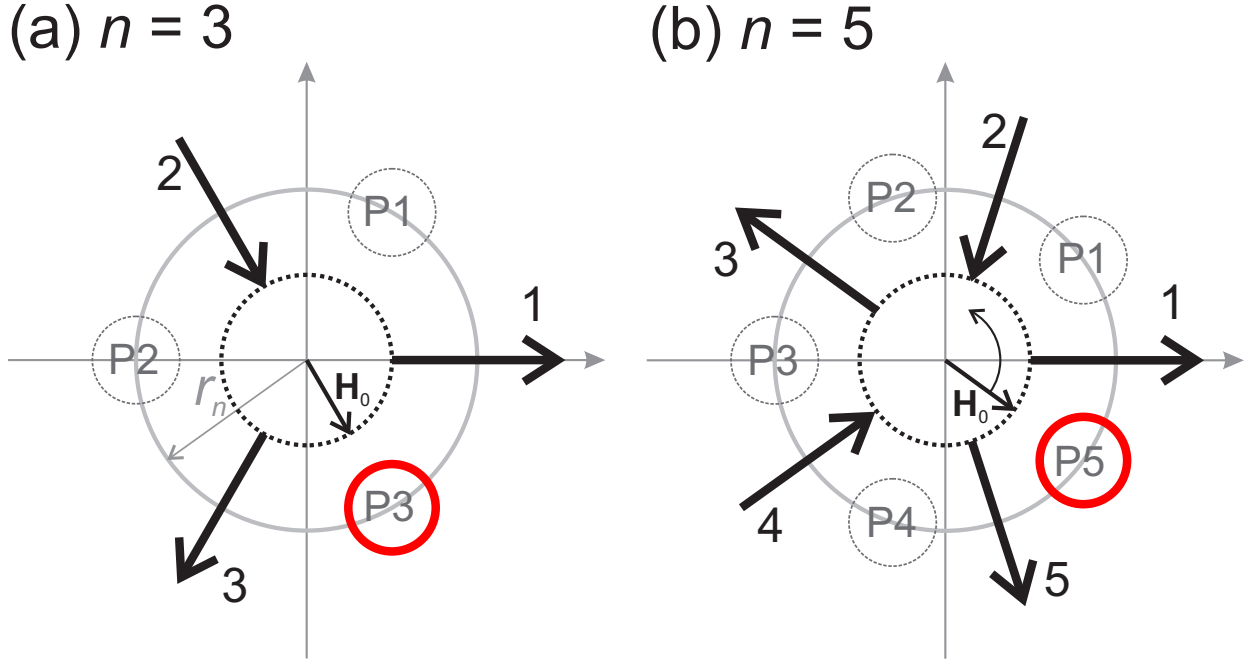}
\end{center}
\caption{
(Color online) Radial odd-$n$ spin ring systems for (a) $n =$~3, and (b) $n =$~5. The ring radius $= r_n$. Spins are number from $i =$~1 to $n$. An intrinsic ground state defect can take positions P$_i$ as indicated by circles. The defect position for the given ground state configuration is emboldened in red at P$_n$. $\textbf{H}_0$ is the initial applied field used in simulations.
\label{fig1}}
\end{figure}

\section{The Model}
The spin ring is illustrated in Fig. \ref{fig1} for $n$ = 3 and 5 spins. The Ising-like spins $\textbf{s}_i$, represented by arrows numbered $i$ = 1 to $n$ anticlockwise, are radially-aligned and equally spaced by an angle $\theta_n = 2\pi/n$ on a ring of radius $r_n$. Spin $i$ experiences a net point-dipolar field from its neighbors $\textbf{s}_j$

\begin{equation}
  \bm{\mathrm{H}}^{\textnormal{d}}_{i}=\sum^n_{i{\neq}j} [3\hat{\bm{\mathrm{x}}}_{ij}(\bm{\mathrm{s}}_j \cdot \hat{\bm{\mathrm{x}}}_{ij})-\bm{\mathrm{s}}_j]/{|\bm{\mathrm{x}}}_{ij}|^{3}
 \label{eqno1}
 \end{equation}

\noindent where \textbf{x}$_{ij}$ is the vector displacement from spin \textbf{s}$_i$ to spin \textbf{s}$_j$. We work in normalized units, and set all $|\textbf{s}_i|$ = 1.

It is instructive to first consider the ground state (GS) of the system. To minimize the dominant pairwise interaction which exists between first nearest neighbors, an ``in-out'' relative configuration must be adopted for a given pair. Further neighbor interactions, whilst not necessarily insignificant, will not alter this ordering rule. Because $n$ is odd, if one attempts to propagate the rule around a ring, a defect must always be ultimately formed, consisting of a frustrated ``in-in'' or ``out-out'' arrangement, resembling a magnetic soliton defect,\cite{2002Cowburn} as illustrated in Fig. \ref{fig1}. The frustrated defect can hence exist at positions P$_i$, as indicated by circles, and we will refer to spins which constitute a soliton as ``defected spins''.  It is important to note that the system can generally support only odd numbers of solitons

Similar solitons are found in linear 1D chains of Ising spins where two domains of opposite phase meet,\cite{2000Cowburn2,2002Cowburn} a result of a two-fold degenerate antiferromagnet-like GS. In our ring system, the GS possesses an \textit{intrinsic} local frustration, much like a 1D Ising chain with periodic boundary conditions where setting $n$ = odd imposes this ``twist'' in the local order parameter. There are also $n$ possible soliton positions. Calculation of the net Zeeman energy  $E = - \frac{1}{2} \Sigma_i{\textbf{s}_i \cdot \textbf{H}^\textnormal{d}_i}$ shows that this is the GS, which is hence 2$n$-fold degenerate. 

An interesting analogy exists here between the spin ring system and the M\"{o}bius loop, a planar strip possessing only one side due a twisted topology.\cite{2004Cador,2007Starostin} Following the antiferromagnetic order parameter around the ring, one finds it must invert once each cycle at the defect, which represents the topological kink of the M\"{o}bius loop. (This analogy would only stand completely true for a spin ring possessing only 1$^{\textnormal{st}}$ nearest neighbor interactions, however, as we will show, this is the dominant interaction defining the system's behavior, hence the same qualitative behavior is expected.)

For now, we simply consider that each spin possesses an intrinsic switching astroid of a given type. As for a linear 1D chain system,\cite{2002Cowburn} it is anticipated that dipolar interactions result in a local instability at a frustrated defect, its two constituent spins being more willing to flip their orientations than non-defected spins, and local stability elsewhere. Flipping an unstable spin, e.g. by applying a suitable magnetic field (not large enough to reverse any stabilized spins), acts only to move the soliton, creating an energetically equivalent GS if the applied field is subsequently removed. However, for identical spins on the linear 1D chain, an applied field cannot preferentially flip one defected spin over the other, due to symmetry, hence a directional ``pressure'' cannot be established in the system. Whilst QD, e.g. in the intrinsic switching fields of the spins, can locally break the symmetry, this produces no net directionality in the system. As can be seen by simple geometrical considerations, the employment of a curved chain, as for the ring system, potentially overcomes both of these issues, by imposing asymmetry when a uniform global applied field is present. It is hence possible to favour the switching of one defected spin over the other, given QD is not too strong. Under a uniform applied field, two defect spins are now generally inequivalent, and their angular offset may be exploited.

A field sequence may hence be applied to propagate the soliton defect as desired. Due to its simplicity in both experiment and simulation, we consider a uniform applied field $\textbf{H}^{\textnormal{a}}$ of constant magnitude, rotating at a constant rate.\cite{2005Allwood,2006Mattheis,2012BudrikisMorgan} Three field regimes must exist. There are two trivial regimes, one in which the applied field magnitude $H^{\textnormal{a}}$ is too small to ever reconfigure the system, and one in which the field is so large that it only acts to polarizes the system. In between there exists a non-trivial field window in which meaningful dynamics should be possible, exploiting the local instability/stability of defected/non-defected spins imparted by the dipolar interactions and the form of the spins' intrinsic switching astroid.

\section{Simulations}
Our simulations are similar to those recently presented in studies of artificial spin ice systems.\cite{2010Budrikis,2012BudrikisMorgan} We set $r_n$ = $n$/3, which keeps the first nearest neighbor interaction approximately constant at $H^{1^{\textnormal{st}}} \approx$~0.1 as a function of $n$ (an approximation which improves as $n$ increases). For a given $n$, the system is primed from a GS configuration, as shown in Fig. \ref{fig1}, with a defect existing at position P$_5$. A field $\textbf{H}^{\textnormal{a}}$ is applied at an initial angle $\theta_{0} = -\theta_{n}/2$, aligned with the initial net moment of the ring, taking field angle $\theta = 0$ to be aligned with spin $\textbf{s}_1$ and the anti-clockwise sense as positive. The field is incremented in anticlockwise angular steps of $d\theta = + \theta_n/24$: this allows for a sufficient angular resolution, and for even division of the $2\pi$ range, maintaining symmetry between $\theta$ and $\theta + \pi$. For a given simulation step, a spin $\textbf{s}_i$ is selected at (pseudo-)random and an attempt is made to flip its orientation. Spin $\textbf{s}_i$ experiences a total field

\begin{equation}
  \bm{\mathrm{H}}^{\textnormal{t}}_{i}=\bm{\mathrm{H}}^{\textnormal{a}} + \bm{\mathrm{H}}^{\textnormal{d}}_{i}.
 \label{eqno2}
 \end{equation}

To represent realistic reversal behavior, a Stoner-Wolfarth (SW) switching criterion is implemented.\cite{1948Stoner} Even for elongated ferromagnetic nanowires which reverse via nucleation and propagation of a domain wall, nucleation often occurs within a coherently rotating sub-volume.\cite{2012OBrien} Spin $\textbf{s}_i$ flips given two inequalities are satisfied:

\begin{equation}
  \pi/2 < \alpha_i < 3\pi/2
 \label{eqno3}
 \end{equation}

\noindent and

\begin{equation}
|{\textbf{H}}^{\textnormal{t}}_{i}| \geq C_i(\sin^{\frac{2}{3}}(\alpha_i) + \cos^{\frac{2}{3}}(\alpha_i))^{-\frac{3}{2}}
 \label{eqno4}
 \end{equation}

\noindent where $\alpha_i$ is the angle between spin $\textbf{s}_i$ and $\textbf{H}^{\textnormal{t}}_i$.  This allows for reversal given the projection of $\textbf{H}^{\textnormal{t}}_i$ onto $\textbf{s}_i$ is antiparallel with $\textbf{s}_i$ (equation (3)), and that $\textbf{H}^{\textnormal{t}}_i$ lies outside the SW astroid of spin $\textbf{s}_i$ (equation (4)). The SW astroid varies between $C_i$ and $C_i/2$ at its maxima and minima respectively, occurring at and halfway between integer multiples of $\pi$/2 respectively, where $C_i$ is a constant. This random selection and test process is repeated until no further spins can flip, upon which the step ends. Note that this is a zero-temperature simulation hence the system only ever makes downward transitions in energy. This is an appropriate starting point when considering nanomagnets which are robustly thermally stable at remanence\cite{2006Wang,2011Morgan,2013Zhang} and which will reverse their magnetization state effectively instantaneously relative to the applied field rotation rate.\cite{2010Budrikis,2012BudrikisMorgan}

\section{Results: Ideal Behavior}
\begin{figure}
\begin{center}
\includegraphics[width=.75\columnwidth]{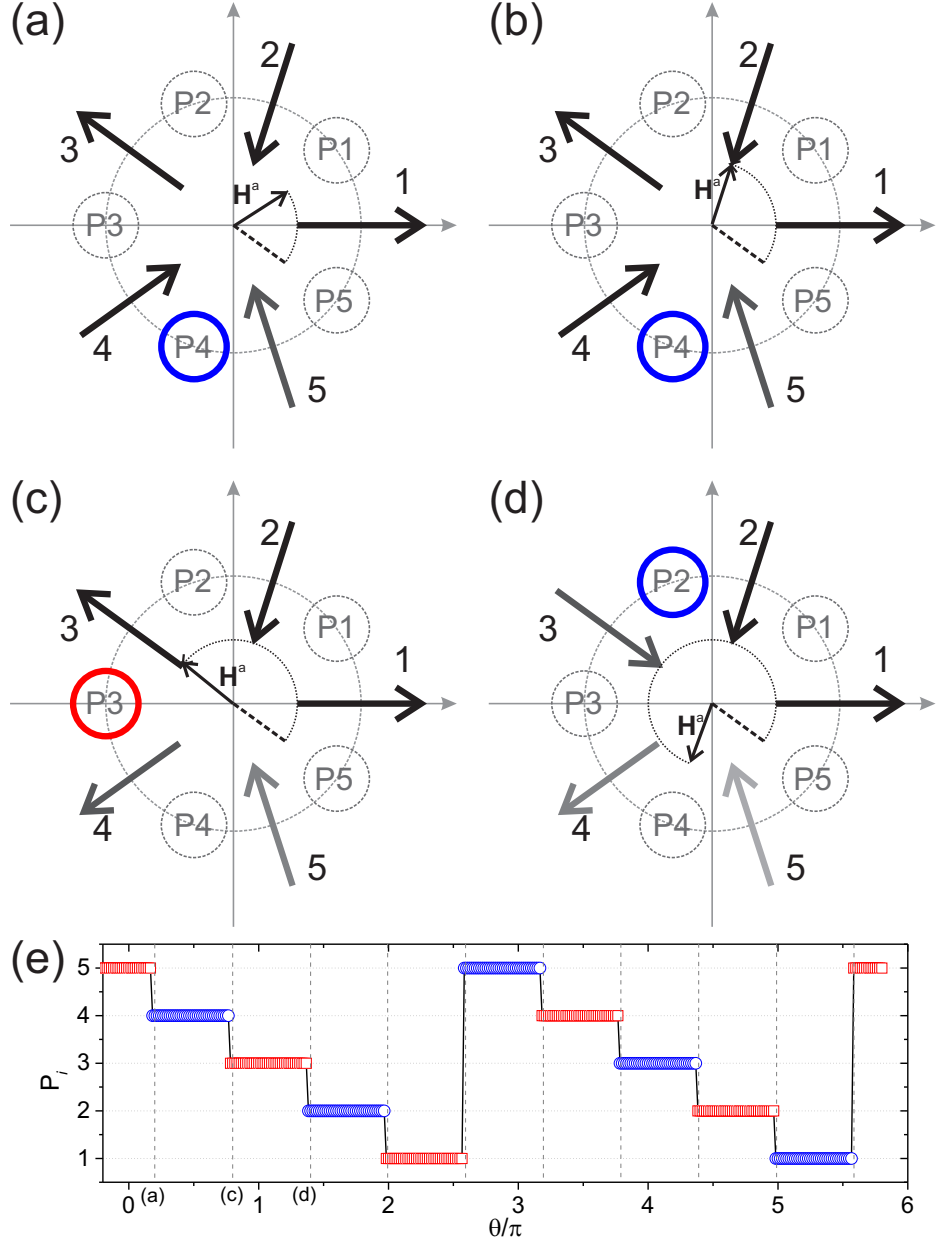}
\end{center}
\caption{
(Color online) (a-d) Ideal soliton propagation for a spin ring with $n =$~5 and $H^\textnormal{a} =$~0.55, as the applied field $\textbf{H}^\textnormal{a}$ is swept anti-clockwise from its initial orientation. $H^\textnormal{a}$ is large enough to sequentially flip defected spins, indicated by lightened gray arrows, but not large enough to polarize the system. The soliton, represented by an emboldened circle, increments its position P$_i$ every $(n - 2)\theta_n/2 = 3\pi/5$ rotation of $\textbf{H}^\textnormal{a}$, as plotted in (e). The red/blue color scheme in (a-e) represents the out/in polarity of a soliton at a given position, as do the square/circular data points in (e). The angular position of the spin flips in (a), (c), and (d) are indicated in (e).
\label{fig2}}
\end{figure}

To illustrate the ideal behavior of the system, we first consider the case in which $C_i =$~1 for all $i$, $n =$~5 and $H^\textnormal{a} =$~0.55.
An animation of a simulation realization is shown in the supplemental material,\cite{2014SM} and we follow the initial behavior schematically in figures \ref{fig1}(b) and \ref{fig2}(a-d). Figure \ref{fig2}(e) shows a plot of the simulated soliton position P$_i$ as a function of applied field angle $\theta$. In the initial configuration, the soliton exists at position P$_5$, as illustrated in Fig. \ref{fig1}(b), with spins $\textbf{s}_{5,1}$ defected.
Whilst $\textbf{H}^\textnormal{a}$ is anti-aligned with $\textbf{s}_3$, $\textbf{s}_3$ is unable to flip, experiencing an opposing net field of 0.3 along its axis ($\ll$~1, at $\alpha_3 = \pi$).
This is true even as $\textbf{H}^\textnormal{a}$ rotates, as $\textbf{H}^{\textnormal{t}}_3$ remains within the SW astroid of spin $\textbf{s}_3$.
As $\textbf{H}^\textnormal{a}$ approaches $\theta \approx \pi/5$, spin $\textbf{s}_5$ is allowed to flip, destabilized by its neighboring spin $\textbf{s}_1$, forming the state shown in Fig. \ref{fig2}(a).
It is worth re-emphasizing that the shape of the SW astroid allows for reversal at such a value of $\alpha_5$.\cite{1948Stoner}
The reversal of $\textbf{s}_5$ propagates the soliton from position P$_5$ to position P$_4$, as shown in Fig. \ref{fig2} (e), acting to stabilize(destabilize) $\textbf{s}_{1(4)}$.
Spins $\textbf{s}_{2,3}$ remain stabilized by their nearest neighbors.
As $\textbf{H}^\textnormal{a}$ continues to rotate, Fig. \ref{fig2} (b), no spin flips occur until $\theta \approx 4\pi/5$ (Fig. \ref{fig2}(c)), at which spin $\textbf{s}_4$ reverses, further incrementing the position of the soliton to position P$_3$.

Continued rotation of $\textbf{H}^\textnormal{a}$ continues to propagate the soliton in this manner (Fig. \ref{fig2}(d,e)) every $(n - 2)\theta_n/2$ rotation of $\textbf{H}^\textnormal{a}$, equal to $3\pi/5$ for $n =$~5. Note that the net ``in/out'' polarity of the defect changes with each increment (indicated by blue/red emboldened circles in figure \ref{fig2} (a-d)), and that the defect position rotates with the opposite sense to that of $\textbf{H}^\textnormal{a}$. Once the defect completes a full circuit of the system, the whole system has undergone a global spin flip transformation from its initial configuration, as the order parameter ``twist'' is swept around. A total rotation of $(n - 2)n\theta_n/2 = (n - 2)\pi = 3\pi$ is required to achieve this. We find the same non-trivial behavior throughout the range of $0.43 \leq H^\textnormal{a} \leq 0.675$, the phase of the spin flipping decreasing(increasing) as $H^\textnormal{a}$ increases(decreases) due to the profile of the SW astroid. Below $H^\textnormal{a} =$~0.43, no dynamics occur as the net field $\textbf{H}^{\textnormal{t}}_i$ is never large enough to satisfy equation (3) for all $i$ and $\alpha_i$. Above $H^\textnormal{a} =$~0.675, the applied field is strong enough to satisfy equation (3) even for non-defected spins, hence the net polarization of the system tracks $\textbf{H}^\textnormal{a}$; whilst multiple soliton defects form under such conditions (for $n > 3$), their motion is trivial, dominated by the Zeeman energy. The non-trivial interval has a width $= \pm H^{1^{\textnormal{st}}} \approx \pm 0.1$ indicating how the operation of the system relies crucially on 1$^{\textnormal{st}}$ nearest neighbor coupling.

We have hence established that such a spin ring system may be used for the manipulation of a well-defined soliton defect. Whilst the system possesses symmetry in its interactions, as for a linear chain, the \textit{anticlockwise} rotating applied field provides the required chirality for unidirectional angular soliton propagation with a \textit{clockwise} sense. It is of course possible to set the initial applied field angle $\theta_0$ to any direction and, in particular, a direction that first favors reversal of spin $\textbf{s}_1$, rather than $\textbf{s}_5$, from the initial spin configuration of Fig. \ref{fig1}(b). In such a case, the soliton will initially take a single step with the \textit{same} sense as $\textbf{H}^\textnormal{a}$ from position P$_5$ to position P$_1$, however, as $\textbf{H}^\textnormal{a}$ continues to rotate the behavior previously described is resumed. Furthermore, reversing the sense of rotation of $\textbf{H}^\textnormal{a}$ to clockwise produces a complementary reversal in the sense of propagation of the soliton, which can be understood by symmetry. Hence, the soliton may be translated to any position in any angular direction.

The scheme also works for alternative switching models and has been tested using an Ising switching astroid. Fundamentally, the scheme requires that spins possess switching astroids which vary as a function of net field direction, that are also offset in angle from each other such that switching can be accessed for individual spins in turn. Hence, the scheme should work experimentally for any realisable bistable nanomagnet type. This is particularly important when considering that the switching astroid of a SW nanomagnet loses four-fold rotational symmetry at finite temperature, the switching threshold for applied fields approaching the easy axis becoming reduced relative to that along the hard axis.\cite{2001Kachkachi} The same qualitative behavior is hence expected.

\section{Results: Disordered Systems}
Next we consider a more realistic situation in which quenched disorder QD is present in the system. We follow similar model studies and implement this as a distribution in the spins' switching behavior.\cite{2011Breitkreutz,2012BudrikisMorgan} For a given realization, we generate the values $C_i$ from a psuedo-Gaussian distribution with mean = 1 and standard deviation $\sigma$. Studies show that this provides a good approximation for the behavior of coupled nanomagnet vertex systems, accounting for a combination of possible property distributions.\cite{2012BudrikisJAP} We explore the behavior of the system as a function of both $H^\textnormal{a}$ and $\sigma$. To further characterize the system, we define the system as ``working'' if the single soliton is able to make a full circuit of the system, as previously described. Over $L$ realizations for each parameter set, we build a map of the probability $p$ that the system operates as designed.

\begin{figure}
\begin{center}
\includegraphics[width=1\columnwidth]{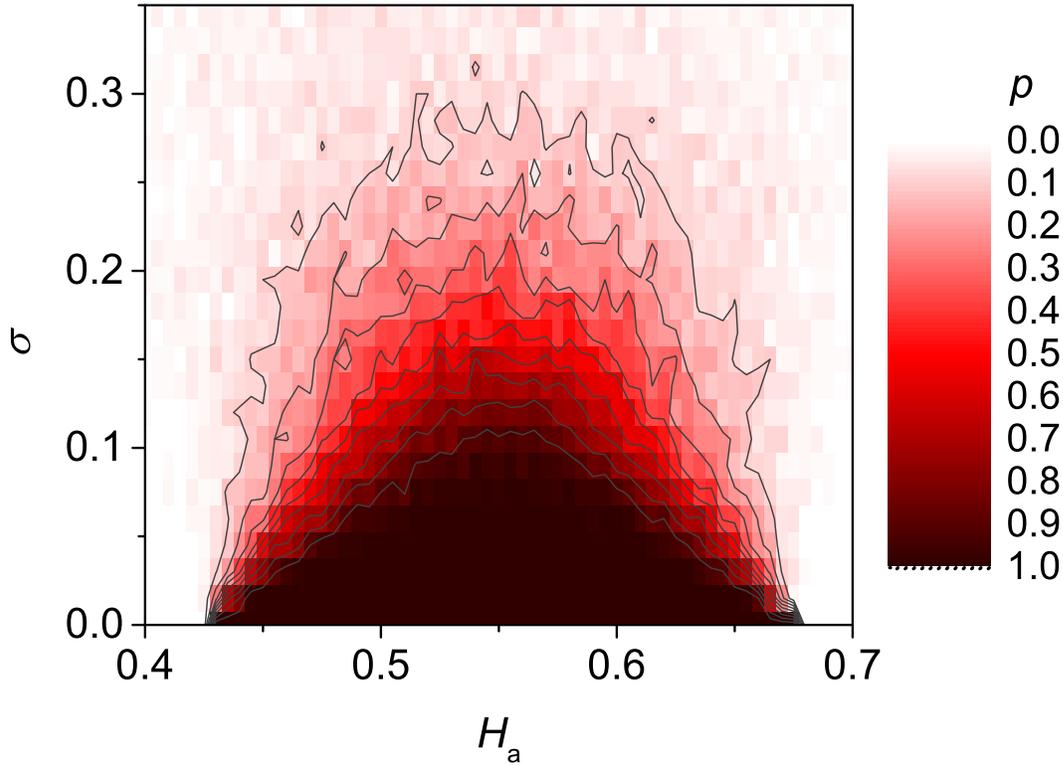}
\end{center}
\caption{
(Color online) Phase diagram of probability of successful operation of an $n =$~5 system, $p$ (see colour key), as a function of applied field magnitude $H^\textnormal{a}$ and switching astroid constant standard deviation $\sigma$.
\label{fig3}}
\end{figure}

\begin{figure}
\begin{center}
\includegraphics[width=1\columnwidth]{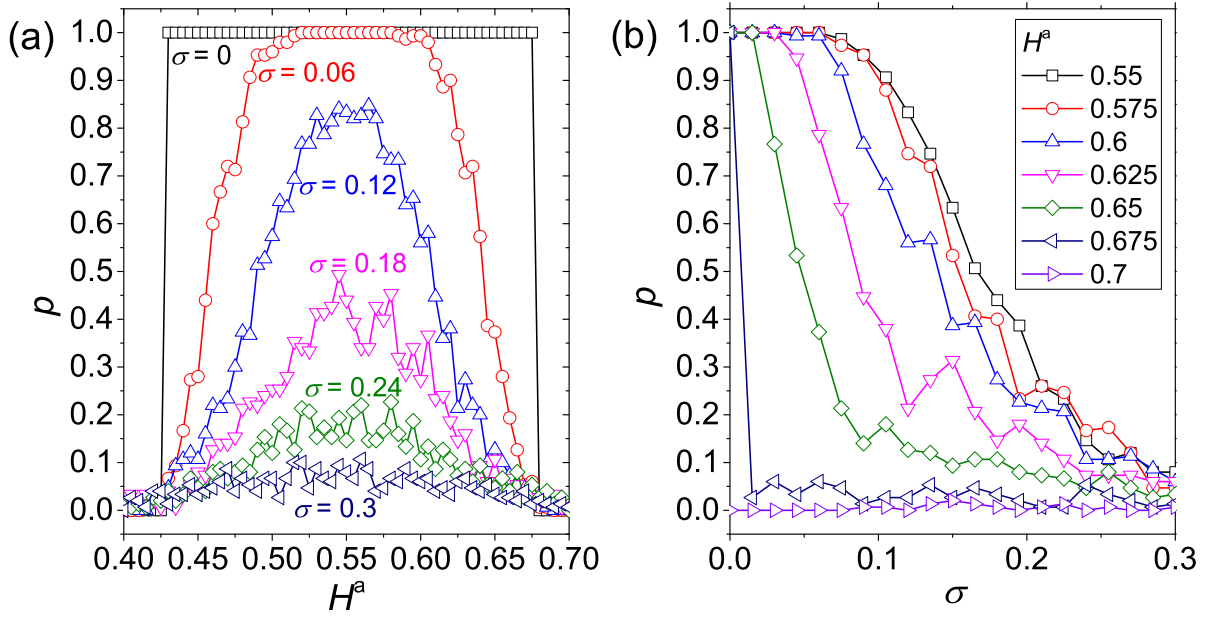}
\end{center}
\caption{
(Color online) Cross-sectional profiles taken through the $p$-map of Fig. \ref{fig3}, taken along the (a) $H^\textnormal{a}$, and (b) $\sigma$ axes at the labeled values of $\sigma$ and $H^\textnormal{a}$ respectively.  Note, due the symmetry of the phase diagram, (b) shows data for $H^\textnormal{a} \geq 0.55$ only.
\label{fig4}}
\end{figure}

This map is shown in Fig. \ref{fig3} for $n =$~5 and $L =$~150. Cross-sectional profiles from Fig. \ref{fig3} are shown in Fig. \ref{fig4} for select values of (a) $H^\textnormal{a}$, and (b) $\sigma$. A clear triangular region with $p \approx$~1 exists spanning the interval $0.43 \leq H^\textnormal{a} \leq 0.675$ at $\sigma =$~0 (as previously discussed), converging linearly to a point at $H^\textnormal{a} \approx$~0.55 and $\sigma \approx$~0.1. $H^\textnormal{a} \approx$ 0.55 is an optimal field magnitude, allowing for the greatest robustness against QD. On moving out of this region, $p$ falls abruptly. The size and shape of this triangular region of $p \approx$~1 is defined by the size of $H^{1^{\textnormal{st}}}\approx$~0.1.

For given values of $H^\textnormal{a}$ and $\sigma$, the behavior within the phase diagram can be understood in terms of spins which are ``pinned'', possessing sufficiently high values of $C_i$ to prevent their reversal even when defected, and spins which are ``loose'', possessing sufficiently low values of $C_i$ such that they can flip even when not defected.\cite{2012BudrikisMorgan} For $H^\textnormal{a} =$~0.45 and $\sigma =$~0.1, there is a significant probability of failure due to at least one spin being unable to flip. Upon meeting a pinned spin, the soliton enters a trapped ``resonant'' behavior mode, taking one step back then one step forward within each rotation of $H^\textnormal{a}$. Failure can also occur if both $\textbf{s}_5$ and $\textbf{s}_1$ are pinned, preventing the soliton from ever moving. As $\sigma$ increases, more erratic behavior can occur, with an increased chance of finding loose spins in a given realization.

For $H^\textnormal{a} =$~0.65 and $\sigma =$~0.1, failure is likely for a given realization due to the increased probability of spins that are loose, which behave trivially under $H^\textnormal{a}$. Typically, a loose spin is allowed to flip when not defected by interactions, nucleating an additional pair of solitons in the system. The evolution of the system then appears as the trivial high field regime discussed in section 3.1. As $\sigma$ increases the probability of such behavior increases. For $\sigma =$~0.4, there often exists increasing numbers of pinned spins too: the combination of loose and pinned spins can result in erratic behavior in which soliton pairs are periodically nucleated and annihilated on the ring, with no meaningful evolution. For $H^\textnormal{a} =$~0.55 and $\sigma =$~0.4, the most erratic behavior is found, due to an average balance of pinned and loose spins, resulting in a balanced probability of various different failure modes. It should be noted that for $\sigma >$~0.4, realizations become increasing unphysical, with negative values of $C_i$ becoming common, hence we do not explore this range.

Within the $p \approx$~1 region in Fig. \ref{fig3}, finite QD acts to modify the number of field steps the soliton spends at each position, which depends on the specific realization, however, given the defect can make a full circuit in the desired sequence, this is still a ``successful'' soliton.

The results shown $(n = 5)$ are representative of all $n > 3$ studied (up to 11), possessing the same form of triangular phase diagram defined by $H^{1^{\textnormal{st}}}$. For $n = 3$, no high-field failure phase is present, as both the trivial high-field regime and non-trivial regime possess a full polarization and a single soliton (Fig. \ref{fig1}(a)). The $n = 3$ system is incompatible with the generation of multiple defects under a uniform applied magnetic field.

Regarding experimental realization, the simulations show that the limit imposed by QD on the system's successful operation is a value of $\sigma \approx H^{1^{\textnormal{st}}}$, for an optimal $H^{\textnormal{a}}$. Keeping within such a limit is in principle experimentally achievable and compares well with recently presented estimates in patterned artificial spin ice systems built from elongated highly-anisotropic bistable NiFe nanomagnets of $\sim$~100~nm dimensions spaced edge-to-edge by $\sim$~40~nm.\cite{2012BudrikisMorgan} QD is however not a straight-forward phenomenon to quantify.\cite{2011Daunheimer}

In order to first generate the required single-defect ground state, it is possible to reset an initial state in an experimental nanomagnet system via various methods currently being developed by the nanomagnetism community. Patterns tailored to allow ``on/off'' switching of thermal dynamics e.g. via volume\cite{2011Morgan,2013Morgan,2013Farhan} or material\cite{2013Zhang,2013Porro}, allow for thermal equilibration of the magnetic macrospins towards their GS, which would remove all but one soliton from any given initial configuration as they undergo a random walk around the ring. It may also be possible to field-anneal the system, using the correct field sequence.\cite{2006Wang,2013MorganFCMP}

\section{Multiturn Counter}
As an example of a practical application, the macrospin ring is a multiturn counter. A full $2\pi$ motion of the soliton requires $(n-2)/2$ complete applied field rotations. As the soliton must travel around the system twice to reset the system, the system can count up to $(n-2)$. Magnetic nano-systems have been shown to be highly applicable for such contactless powerless operation.\cite{2006Mattheis,2007Diegel,2009Diegel,2012Mattheis} This spin ring can potentially be realized experimentally by patterning of radially-aligned single-domain nanomagnets, as discussed, which experience the field of a rotating permanent magnet. Each GS soliton state represents a unique macrospin configuration, which could be directly read by incorporation of Giant Magnetoresistance-based sensing. There is no need to inject soliton defects as for spiral domain wall conduit counters,\cite{2006Mattheis,2009Diegel} which also require a fixed rotational sense to operate and are limited to a maximum number of turns, domain walls eventually ``falling out'' of the ends. The spin ring will count up to $n$ cyclically: each soliton step counts an angle $(n-2)/2n$, which = $3/10$ for $n = 5$, and converges to $1/2$ as $n$ increases.

The spin ring bares similarity to closed-loop conduit devices\cite{2012Mattheis} and perpendicular magnetic anisotropy shift register loops\cite{2013Kiermaier} recently presented, always possessing at least one local defect, which may be used to count field oscillations. A spin ring system built of highly anisotropic  single-domain nanomagnets may also present further benefits, minimising switching time between states relative to extended nanowires.

Furthermore, combining $u$ spin rings of different spin number $n_u$ allows for a total of $\Pi_u(n_u - 2)$ rotations to be counted in a coprime scheme.

\section{Discussion}
The odd-numbered macrospin ring demonstrates an unexplored means of imposing chirality in systems of coupled single-domain nanomagnet chains, as utilized in conventional magnetic logic and MQCA architectures.\cite{2000Cowburn2,2003Imre} The scheme is simple to implement using nanopatterned thin films and more general systems of curved spin chains may be designed. It is further an example of a user-designed artifical geometrically frustrated system, possessing an \textit{intrinsic} soliton defect, which can be manipulated in a well-defined way and potentially employed for useful operations, exemplified here as a simple multiturn counter. It further represents a building block for the study of more complex coupled systems.\cite{2013Bhat} Whilst a small handful of reports exists in the field of molecular magnetism on such ``magnetic M\"{o}bius loops'',\cite{2004Cador,2012Baker} built from odd-numbered spin-$1/2$ rings, our work highlights the possibility of exploring such physics in nanopatterned systems via real-space real-time microscopy.\cite{2006Wang,2013Farhan}

\begin{acknowledgments}
We thank C. H. Marrows, Z. Budrikis, R. Mansell, and A. Fernandez-Pacheco for useful discussions and careful proof-readings. We acknowledge research funding from the (UK) EPSRC, and the European Community under the Seventh Framework Programme Contract No. 247368: 3SPIN. This work has been partly carried out within the Joint Research Project EXL04 (SpinCal), funded by the European Metrology Research Programme. The EMRP is jointly funded by the EMRP participating countries within EURAMET and the European Union.

\end{acknowledgments}

\end{document}